\documentclass[aps,prb,singlecolumn,superscriptaddress]{revtex4}
\usepackage{graphicx}
\newcommand\beq{\begin{equation}}
\newcommand\eeq{\end{equation}}
\newcommand\bear{\begin{eqnarray}}
\newcommand\eear{\end{eqnarray}}
\begin{document}
\title{Interplay between Bonding and Magnetism in the Adsorption of NO on Rh Clusters}
\author{Prasenjit Ghosh}
\affiliation{Theoretical Sciences Unit, Jawaharlal Nehru Centre 
for Advanced Scientific Research,\\
 Jakkur, Bangalore 560 064, India }
\author{Raghani Pushpa}
\affiliation{International School for Advanced Studies (SISSA), via Beirut 2-4,
Trieste 34014,Italy}
\author{Stefano de Gironcoli}
\affiliation{International School for Advanced Studies (SISSA), via Beirut 2-4,
Trieste 34014,Italy}
\affiliation{CNR-INFM DEMOCRITOS National Simulation Center, via Beirut 2-4, Trieste 34014, Italy}
\author{Shobhana Narasimhan}
\affiliation{Theoretical Sciences Unit, Jawaharlal Nehru Centre 
for Advanced Scientific Research,\\
 Jakkur, Bangalore 560 064, India }
\date{\today}
\baselineskip 24pt
\begin{abstract}
\baselineskip 24pt
We have studied the adsorption of NO on small Rh clusters, containing one to five
atoms, using density functional theory in both spin-polarized and non-spin-polarized
forms. We find that NO bonds more strongly to Rh clusters than it does to Rh(100) or
Rh(111); however, it also quenches the magnetism of the clusters. This (local) effect
results in reducing the magnitude of the adsorption energy, and also washes out the
clear size-dependent trend observed in the non-magnetic case. Our results illustrate
the competition present between the tendencies to bond and to magnetize, in small clusters.

\end{abstract}

\maketitle
\newpage

\section{INTRODUCTION}
\label{intro}
The catalytic
dissociation of NO is one of the important rate-limiting steps in the conversion
of automobile exhaust gases into less undesirable products. Many precious metals are known to facilitate the catalytic dissociation of NO; of these,
Rh appears to be the most efficient catalyst, i.e., the energetic barrier that
must be overcome before the N-O bond can be broken has been found to be the lowest
on Rh surfaces.\cite{Loffreda1} In general, such barriers tend to decrease
on decreasing the coordination of the metal (catalyst) atoms. For 
example, experiments\cite{Borg, Hopstaken, Villarubia} have shown that the barrier
for NO dissociation is lowered from 0.67 eV to 0.38--0.46 eV on going from the Rh(111)
surface, where surface atoms have ninefold coordination, to the Rh(100) surface, where atoms at the surface have eightfold coordination. Similarly, Ertl and collaborators have shown that NO molecules on the Ru(0001) surface dissociate preferentially at step edges, where atoms are less coordinated than on flat terraces.\cite{Zambelli} In confirmation of
their experimental observation, Hammer has shown\cite{Hammer} that the dissociation barrier for NO
is lowered drastically, from 1.28 eV to 0.15 eV, upon moving from terrace atoms to step
edge atoms. (However, there are exceptions to this trend, e.g., stepped Rh(100) surfaces,
despite possessing  lower coordination than flat Rh(100), display a higher barrier
for NO reduction.\cite{Loffreda1})
This raises the question of whether small
Rh nanoparticles, which may be expected to have a very high surface-to-volume
ratio, and thus a high proportion of under-coordinated sites, may lower the dissociation
barrier even further, relative to the value on Rh(100). 

While Rh clusters thus constitute promising candidates for the nanocatalytic
dissociation of NO, there is however another factor to be considered: small Rh clusters are unusual
in that they are magnetic, even though bulk rhodium is non-magnetic.\cite{Reddy1, Cox1, Cox2} At first
sight, magnetism might be expected to further enhance the catalytic activity of small Rh clusters: 
due to the
familiar phenomenon that magnetism tends to increase interatomic distances, magnetic clusters
will possess a lower effective coordination, and thus conceivably be better catalysts, than non-magnetic ones. However, as we will show below, the situation is somewhat more complicated than this.
Thus, while the presence of magnetism makes
issues regarding Rh nanocatalysis more complex, these systems offer a good opportunity for
studying the interplay between magnetism and coordination number in the efficacy  of
magnetic catalysts. 

As a first step towards addressing such issues, in this paper we
present results from a density functional theory (DFT) study of the adsorption of NO on very small Rh
clusters, where the number of atoms $n \leq 5$. These Rh$_n$NO complexes constitute the
initial state for the rate-limiting step in the catalytic reduction of NO. Moreover, the strength of the adsorption may be expected to
give some indications about how easy it is to dissociate NO on the Rh$_n$ cluster,
the argument being that a greater adsorption energy would indicate stronger bonding
between the metal atoms and NO, which in turn may signal a weakening of the N-O bond,
thus facilitating dissociation.

The main motivation of our study is to see how size affects
the bonding ability of Rh clusters. Recent work on clusters has shown that the study
of the size-dependence of their properties falls naturally into two size regimes. At
larger sizes (hundreds to thousands of atoms), the properties evolve smoothly towards
those of the bulk; whether they approach the bulk properties from above or from below,
depends on the property under examination, as well as, in some cases, the element
under consideration. However, at smaller cluster sizes, one frequently observes trends that
are non-monotonic, so that a specific size of cluster may possess desired properties,
while adding or removing even a single atom may alter the properties drastically. 
Such oscillations can, however, make more apparent the factors governing the behavior
of clusters, thus enabling one to gain insight into even the properties of larger clusters.

As an example of a similar system whose nanocatalytic properties have attracted recent attention, 
we mention Au clusters (with 20 or less atoms) supported on magnesia, which have been found, both experimentally and
theoretically\cite{Sanchez, Hannu, Yoon, Walter, Heiz} to facilitate the oxidation of
CO,  even though bulk Au and Au surfaces are, famously, chemically inert. Similarly, the
reaction rates of the hydrogenation of toluene\cite{Xu} and the cyclomerization of
acetylene\cite{Abbet} on various clusters (in some cases supported on oxide substrates) have also been found to vary sensitively as a function of
cluster size.

The rest of this paper is structured as follows: In Section \ref{Prev} we summarize previous
experimental and theoretical work, both on bare Rh clusters and Rh$_n$NO complexes. In
Section \ref{FPdet} we present some of the technical details of our calculational method.
In Section \ref{results} we present our results, first on the bare Rh clusters, and then on
the clusters with NO adsorbed on them. In both cases, we present separately the results of
spin-polarized and non-spin-polarized calculations. Finally, in Section \ref{Disc} we summarize
and analyze
our results and discuss some of their implications. 

\section{Previous work on Rh$_n$ and NO-Rh$_n$}
\label{Prev}
The possibility that Rh clusters (unlike bulk Rh) may be magnetic was first
suggested by  the DFT calculations of Reddy {\it et al.}\cite{Reddy1} In their work,  thirteen-atom clusters of Pd, Rh and Ru were all found to have non-zero magnetic moments, with the largest value (of
21 $\mu_B$, corresponding to 1.61 $\mu_B$ per atom) being displayed by Rh$_{13}$. They attributed this to the reduced dimensionality
as well as the high symmetry of the icosahedral thirteen-atom clusters. Subsequently,
other authors have found other geometries for Rh$_{13}$ that were lower in energy, with
a lesser (but non-zero) magnetic moment.\cite{Bae1, Bae2, Chang, Rogan} Extending this work to
other sizes, several
groups have performed calculations to determine the structure, binding energies and
magnetic properties of small Rh clusters using DFT as well as other quantum chemical techniques.\cite{Futschek, Reddy2, Nayak, Chien, Bae1, Endou1, Jinlong, Harding} There is considerable variation in the structures, binding energies and spin multiplicities obtained
by various authors.
For example, while many authors\cite{Nayak,  Reddy2, Futschek, Chien} predict that the lowest-energy structure of Rh$_4$ is non-magnetic with a tetrahedral geometry, other calculations\cite{Endou1} suggest
that a spin septet state, also in a tetrahedral geometry, is favored. 
Similar issues exist also for Rh$_5$ and Rh$_6$.

Experimental studies on bare Rh clusters have been carried out by Cox {\it et al.}\cite{Cox1,
Cox2} who found that the clusters do indeed possess rather large magnetic moments,
ranging from 0.3 -- 1.1 $\mu_B$ per atom; the magnetic moment per atom decreases with the size
of the cluster, becoming zero in the neighborhood of $n = 60$. However, the smallest
cluster studied by them corresponds to $n = 8$. The only experimental work on smaller
bare Rh clusters that we are aware of is that of Gingerich {\it et al.} who studied Rh$_2$,\cite{Gingerich}
and determined its binding energy and bond length.

The adsorption of NO on a Rh dimer has been examined theoretically in a DFT study by Endou
{\textit et al.};\cite{Endou2} they found that the N-O bond length is elongated (with respect to the gas
phase) due to a back-donation mechanism whereby electrons are transferred from the dimer
to the adsorbate. The same group also performed a comparative study of NO adsorption on four-atom
clusters of Rh, Pd, Ag, Ir, Pt and Au, finding that the adsorption is most favored on Ir$_4$.\cite{Endou1}

There have also been a few experimental studies of NO adsorption and dissociation on small Rh clusters.
In their Fourier transform ion
cyclotron resonance studies of reactions of NO with Rh$_6^+$,
Ford \textit{et al.}\cite{Ford} observed biexponential kinetics,
which they have interpreted in terms of structural isomerism.
In another experimental study of NO decomposition on small, charged Rh clusters,
Anderson \textit{et al.}\cite{Anderson}
have reported that, for both cationic and anionic clusters, the
reaction rate increases smoothly with cluster size, though the reaction
proceeds significantly faster on cationic clusters than on 
anionic ones. Motivated by these experimental studies, Harding \textit{et al.}\cite{Harding} 
have performed DFT studies of NO adsorption and dissociation
on structural and spin isomers of Rh$_6^+$. Their studies suggest
that the biexponential kinetics observed by Ford \textit{et al.} is indeed
due to the presence of structural isomers of Rh$_6^+$, rather than due
to different spin states. They found that the energy barrier for NO dissociation
on Rh$_6^+$, with a geometry of a trigonal prism, is markedly lower 
(0.23-0.36 eV depending on the spin
state) than that reported
for the same reaction on the Rh(100) surface ($\sim$ 0.5 eV\cite{Loffreda3}).

\section{DETAILS OF \textit{AB INITIO} CALCULATIONS} 
\label{FPdet}

All our DFT calculations have been performed using the PWscf code, which forms a part of
the Quantum-ESPRESSO distribution.\cite{qnesp} The Kohn-Sham equations\cite{Kohnsham} were
expanded in a plane-wave basis set with a cut-off of 30 Ry, while a larger cut-off of 216 Ry
was used for the augmentation charges introduced by the ultrasoft pseudopotentials\cite{Vanderbilt}
that we used to describe the electron-ion interactions. Since we wanted to investigate the
effects of magnetization, all our calculations were performed using both the spin-polarized (SP)
and non-spin-polarized (NSP) versions of the Perdew-Burke-Ernzerhof (PBE)\cite{GGA4} form of the generalized gradient approximation (GGA). Since the code makes use of periodic boundary conditions,
the clusters were placed in a box of side 12 \AA; this size is large enough to ensure that the
interaction between periodic images is negligible. Accordingly, Brillouin zone integrations were
performed using only the $\Gamma$ point. In order to hasten convergence to self consistency, we 
have used a very small Gaussian smearing with a width of 0.002 Ry -- note that it is important that
this smearing be small, since we have found that larger values of the smearing width can lead to 
errors in the magnetic moment of the lowest-energy configuration.

As a reference, we have performed calculations on bulk Rh and NO in the gas phase.
For the former, we obtain a lattice constant of 3.86 \AA\ and a bulk modulus
of 254 GPa, which are
in good agreement with the experimental values of 3.80 \AA\ \cite{Rhbexpt} and 269 GPa,\cite{Rhbexpt} while for the
latter we obtain an N-O bond length of 1.17 \AA. This too closely matches the
experimentally determined value of 1.15 \AA. \cite{Johnson}  

In agreement with previous work, we have found that the energy landscapes of the Rh$_n$ and
Rh$_n$NO complexes possess a great many nearly degenerate local minima, both in
coordinate space and in spin space. For this reason, we have made use of a very large number
of starting configurations (both geometry and spin), and also performed some calculations where
the magnetic moment was constrained. We are therefore reasonably confident that we have found
the global-minimum structures. Structural optimization was performed using 
Hellmann-Feynman forces \cite{Hellmann, Feynman} and a BFGS-based
algorithm \cite{BFGS} for the minimization of energy. Moreover, no symmetry constraints
were imposed when performing structural optimization, so as to ensure that distortions were permitted.

For Rh$_1$ and Rh$_2$, the question of choosing initial geometric configurations
is trivial. 
For the bare Rh$_3$ cluster, we tried different triangle-based geometries
[equilateral (eq), isosceles (isos) and scalene] as starting configurations.
For the initial geometries for Rh$_4$ and Rh$_5$, we used the lowest-lying structural isomers
reported in the literature:\cite{Nayak, Reddy2, Futschek} the square (sq) and the tetrahedron 
(tet) geometry for Rh$_4$ and the triangular bipyramid (tbp) and the square
pyramid (sqp) for Rh$_5$. Further, we made use of a variety of starting spin states; in a few cases we found that it was necessary to constrain the value of the magnetization in order to
find some low-lying states.

The situation becomes considerably more complex when considering NO adsorption on the 
Rh$_n$ clusters, since there are a large number of inequivalent adsorption sites, as well
as degrees of freedom corresponding to the orientation of the NO molecule relative to
the cluster. We have therefore considered a great many starting geometries, and in
the majority of cases we found that these relaxed to different local minima of the Rh$_n$NO
complex. The number of inequivalent possibilities that have to be considered
increases rapidly as the size of the cluster grows, and it is therefore a rather
challenging task to find the global
minimum structure. 

\section{RESULTS}
\label{results} 

Though it is well-established that small Rh clusters are magnetic, we have performed
our calculations both permitting spin polarization (SP) and suppressing it (NSP). A
comparison of the SP and NSP calculations should shed some light on the consequences
of magnetism. Our results are grouped below into four sub-sections: (A) bare clusters + SP,
(B) bare clusters + NSP, (C) Rh$_n$NO + SP, and (D) Rh$_n$NO + NSP.

\subsection{Bare Rh clusters: Spin-Polarized}
\label{baresp}

For a single Rh atom, we obtain a ground state that agrees with that which is well-established, both
theoretically and experimentally: a $^4$F (4d$^8$5s$^1$) state,
with five spin-up ($\uparrow$) and three spin-down ($\downarrow$) electrons in the 4d
orbital and one $\uparrow$ electron in the 5s orbital. This corresponds to a magnetic moment of
3 $\mu_B$.

For the Rh dimer, we obtain a bond length of 2.25 \AA\ and a binding energy (BE) of 1.48 eV.
The former is in excellent agreement with experiment\cite{Gingerich} and previous 
calculations,\cite{Nayak} while the latter closely matches the experimentally reported value
of 1.46 eV,\cite{Gingerich} as well as some theoretical values.\cite{Nayak} We note that the calculated
values for the BE reported in the literature vary over a range, depending on the level of theory used.
We obtain a magnetic moment of 2 $\mu_B$/atom, corresponding to a spin multiplicity ($2S+1$, where $S$ is the total spin of the cluster) of 5. This too is in agreement with earlier experiments and calculations.

There is disagreement in the literature about whether the ground state of Rh$_3$ is an equilateral
triangle with a spin multiplicity of 4,\cite{Nayak} or an isosceles triangle with a spin multiplicity
of 6.\cite{Reddy2} The lowest energy configuration found by us is in agreement with the latter. The
next-lowest-lying isomer found by us is indeed an equilateral triangle, but its spin multiplicity is
6 and not 4. These two lowest-lying isomers are separated by only 0.008 eV/atom in BE.
The equilateral triangle in the quartet spin state lies still higher, by an amount of 0.04 eV/atom.

Low-lying isomers of Rh$_4$ have either a square or tetrahedral geometry. The lowest-energy configuration
found by us corresponds to a tetrahedron with a spin multiplicity of 7; this is in agreement with
one set of earlier calculations.\cite{Endou1} We find two energetically degenerate isomers that lie
higher than this by an amount of 0.05 eV/atom: a square geometry with a spin multiplicity of 5, and
a non-magnetic tetrahedron. We note that some previous studies\cite{Nayak, Reddy1, Futschek} have claimed
that the latter configuration corresponds to the lowest-energy isomer.

We find two degenerate lowest-energy configurations for Rh$_5$: both are square pyramids, one with spin multiplicity of 6, and the other with spin multiplicity of 8. In earlier work, one set of authors had
found the former to be the lowest-lying isomer,\cite{Futschek} while another had found the latter.\cite{Reddy2} We find that these lie lower than a triangular bipyramid with $2S+1=8$ by a small amount of 0.03 eV/atom. 

From these results, one can see that even at these very small cluster sizes, there are a large number
of nearly degenerate spin and structural isomers, and it seems likely that several isomers will be
simultaneously present upon experimentally preparing Rh clusters.

As mentioned earlier, we are interested in seeing what effect the coordination number has on reactivity.
The nominal coordination number of the lowest-lying isomer increases with the size of the cluster, having values of 0, 1, 2, 3 and 3.2 as $n$, the number of atoms, is increased from 1 to 5. As expected,
as the nominal coordination increases, the interatomic bond lengths increases. Since the effects on electronic structure of increased coordination number and longer bond lengths are expected to be linked
and correlated, it is useful to define a quantity that simultaneously incorporates both effects. We therefore define the effective coordination number of the $i^{\rm th}$ atom in a cluster by:\cite{corelev}

\beq
N_{eff}(i)=\frac{\sum_{j\ne i}\rho_{Rh}^{at}(R_{ij})}{\rho_{Rh}^{at}(R_{bulk})},
\eeq

\noindent where the sum is calculated over all the other atoms $j$ in the cluster,
$R_{ij}$ is the distance between atoms $i$ and $j$, $\rho_{Rh}^{at}(R)$ 
is the computed spherical charge density distribution of an isolated Rh atom 
at a distance $R$ from the nucleus, and $R_{bulk}$ is the nearest-neighbor
bond length in the bulk. In other words, $N_{eff}(i)$ contains information about the ambient
electronic density (due to the other atoms) that the atom $i$ is embedded into. This is
in the spirit of the embedded-atom method\cite{Daw} or effective-medium-theory,\cite{EMT}
making the approximation that the density due to the neighboring atoms can be approximated
by the sum of the atomic densities. The average effective coordination number of a cluster
is then given by $\langle N_{eff} \rangle = (1/n) \sum_i N_{eff}(i)$.

\begin{figure}[p]
\centering
\includegraphics[scale=0.40, angle=-90]{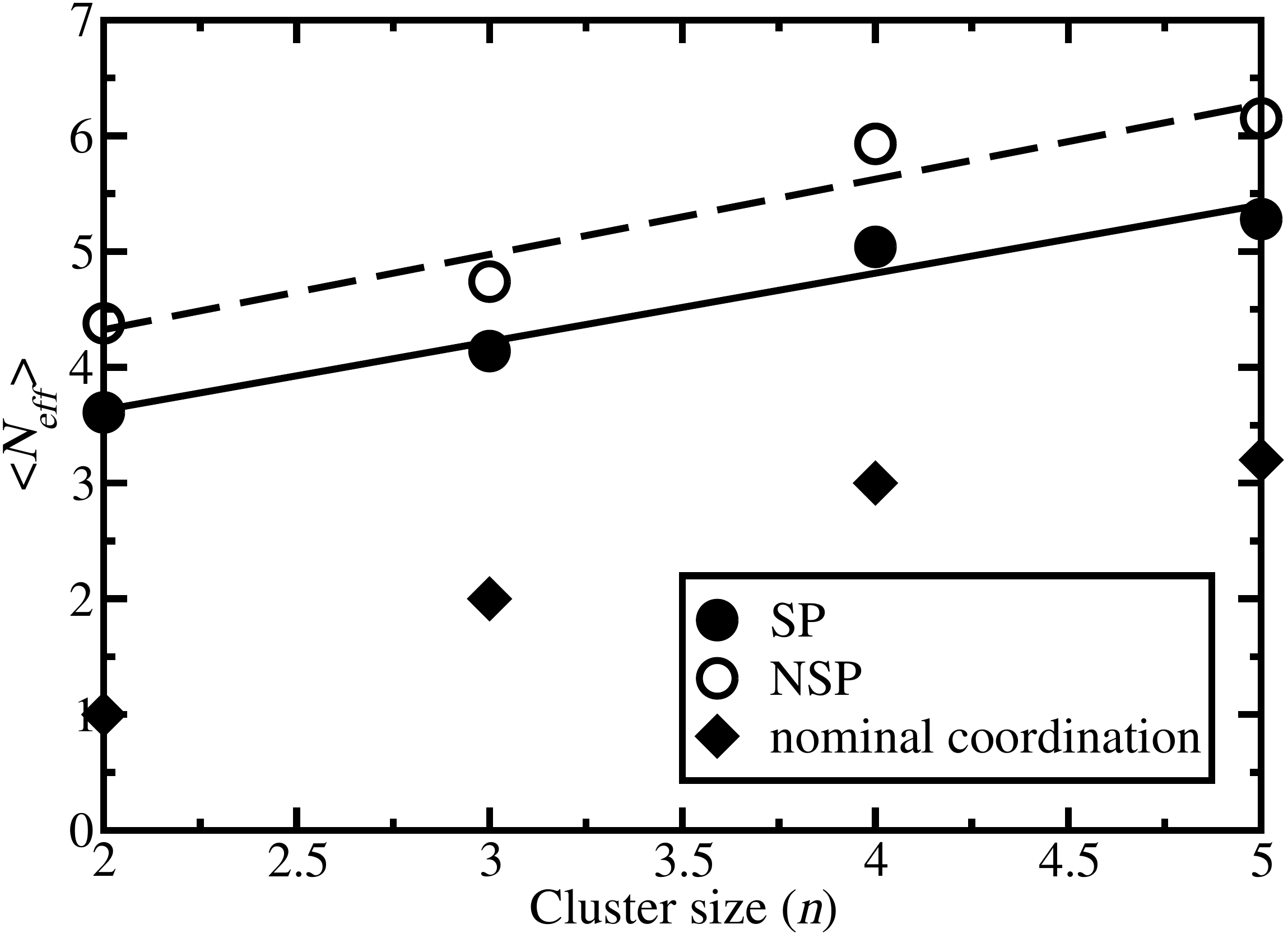}
\caption{Effective coordination number ($<N_{eff}>$) vs. cluster size (n).
The filled circles and open circles denote the $<N_{eff}>$ of the clusters
for SP and NSP cases respectively. The filled diamonds show the nominal
coordination in each cluster. The solid and the dashed straight lines are
guides to the eye for SP and NSP respectively.}
\label{fig_neff}
\end{figure}

The filled black circles in Fig.~\ref{fig_neff} show how $\langle N_{eff} \rangle$ varies with
$n$ for the lowest-lying isomers. It can be seen that the variation is approximately linear,
and that $\langle N_{eff} \rangle$ is significantly larger than the nominal coordination (filled
black diamonds). 

\begin{figure}[p]
\centering
\includegraphics[scale=0.40, angle=-90]{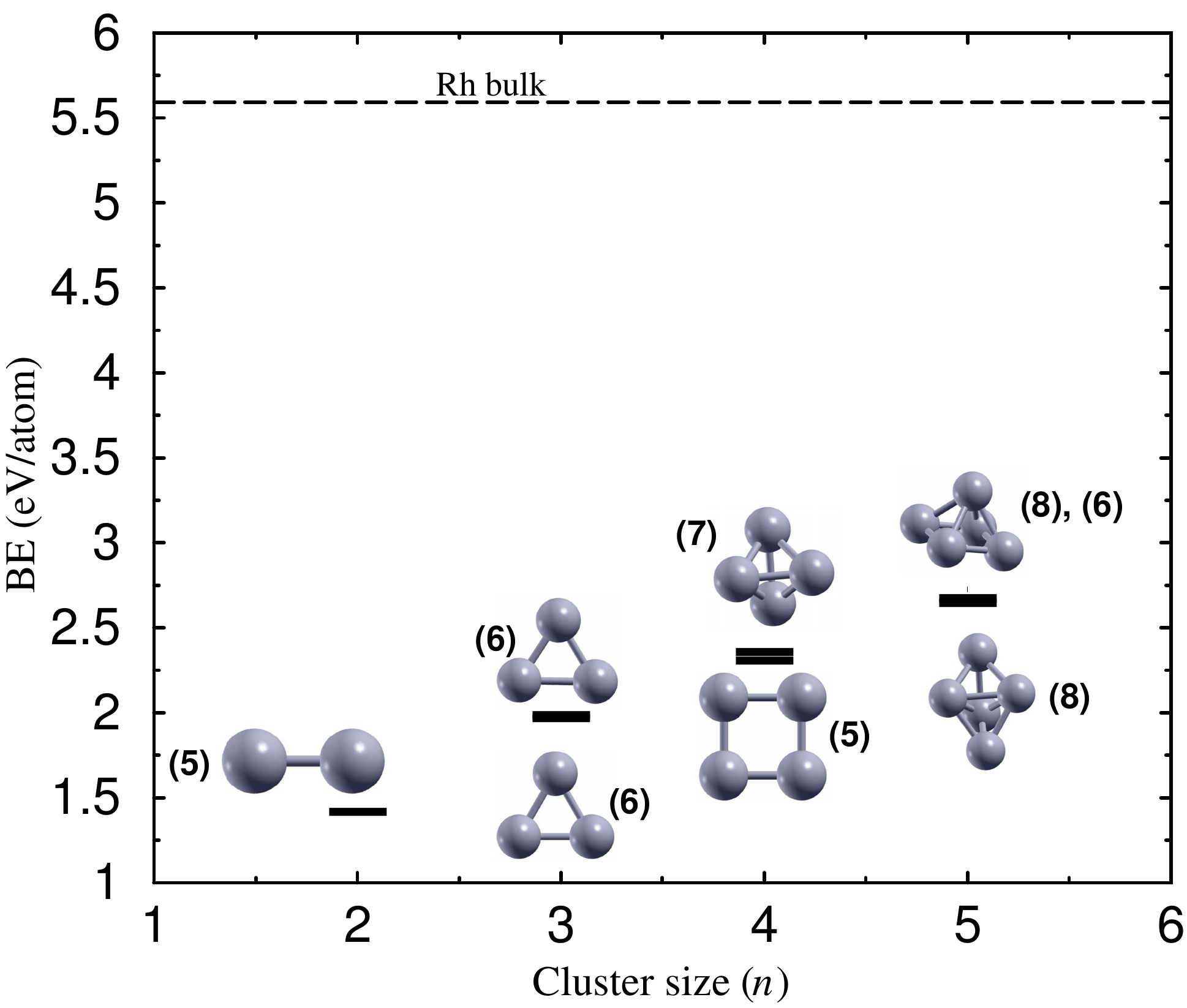}
\caption{Binding energies vs. ~cluster size (for SP). The horizontal line
corresponds to the BE per Rh atom in the bulk. The equilibrium structures of
different clusters have been drawn with the spin multiplicity
given in parentheses.}
\label{fig_ebin_sp}
\end{figure}

\begin{figure}[p]
\centering
\includegraphics[scale=0.40, angle=-90]{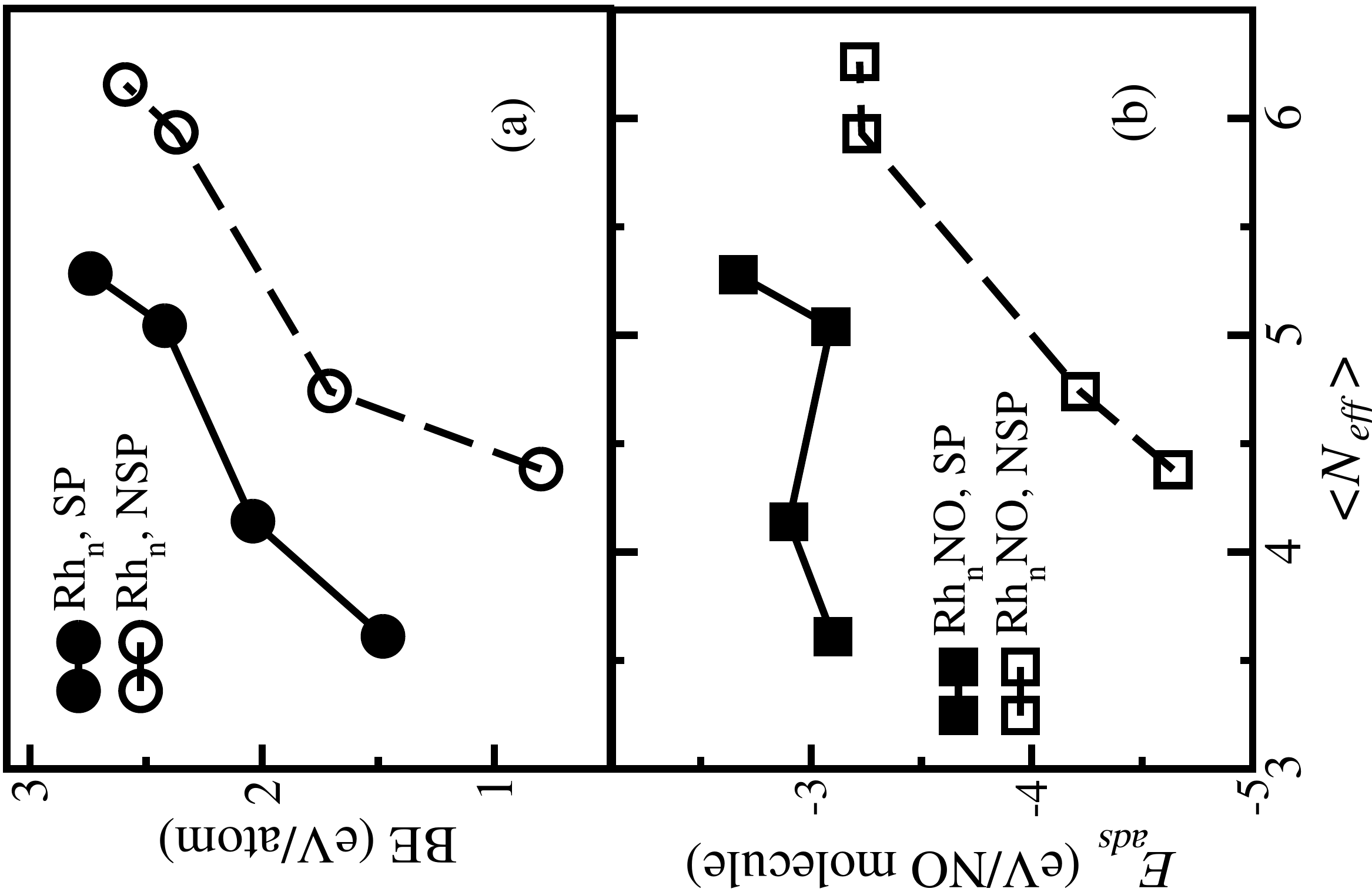}
\caption{(a) Binding energy for the bare clusters and (b) NO adsorption energy
vs. ~effective coordination number.
The filled and open circles show the BEs of Rh$_n$ for SP and NSP
respectively. The filled squares and the open squares denote
$E_{ads}$ of Rh$_n$NO complexes in SP and NSP cases respectively.}
\label{fig_be+eadsvsneff}
\end{figure}

Our results for the bare Rh clusters are summarized in Table~\ref{tab_bare_sp}, where we have
also compared our results to those of previous calculations, while in Fig.~\ref{fig_ebin_sp}
we show the BE and structures of low-lying isomers. Note that both the spin multiplicity
and the BE/atom of the lowest-lying isomer also increase monotonically with $n$, and that
the energy difference between the two lowest-lying isomers is indeed minute in all cases.
However, we note that the magnetic moment {\it per atom} decreases monotonically with increasing
$n$, which is in agreement with the expectation that larger clusters, being more highly coordinated,
should display a decreased tendency towards magnetization.
The filled black circles in Fig.~\ref{fig_be+eadsvsneff}(a) show that the binding energy per
atom varies more-or-less smoothly also with $\langle N_{eff} \rangle$.

\subsection{Bare Rh clusters: Non-Spin-Polarized}
\label{barensp}

When we repeat our calculations upon constraining the clusters to be non-magnetic, it becomes
immediately obvious that the suppression of magnetism has a noticeable impact on structure. Our
results are summarized in Table~\ref{tab_bare_nsp}. In
agreement with general experience, one finds that the clusters contract upon performing NSP
calculations. This is because in the SP case, magnetism (which is favored by Hund's
rule) is in competition with the tendency to form interatomic bonds; upon suppressing the former,
the latter tendency is increased, resulting in shorter interatomic bonds and hence larger values
of $\langle N_{eff} \rangle$ (see the open circles in Fig.~\ref{fig_neff}). While for the SP case,
both isosceles and equilateral triangles constitute stable geometries for Rh$_3$, in the NSP situation,
the former relaxes to the latter. We note also that the
near-degeneracy between the square and tetrahedral structures of Rh$_4$ appears to be lifted on
suppressing spin polarization.

The open circles in Fig.~\ref{fig_be+eadsvsneff}(a) show that the BE varies monotonically with $\langle N_{eff} \rangle$ also
when magnetism is suppressed; this figure makes it clear that allowing magnetism stabilizes the clusters,
while decreasing their effective coordination.

\subsection{NO on Rh clusters: Spin-Polarized}

\begin{figure}[p]
\centering
\includegraphics[scale=0.40, angle=-90]{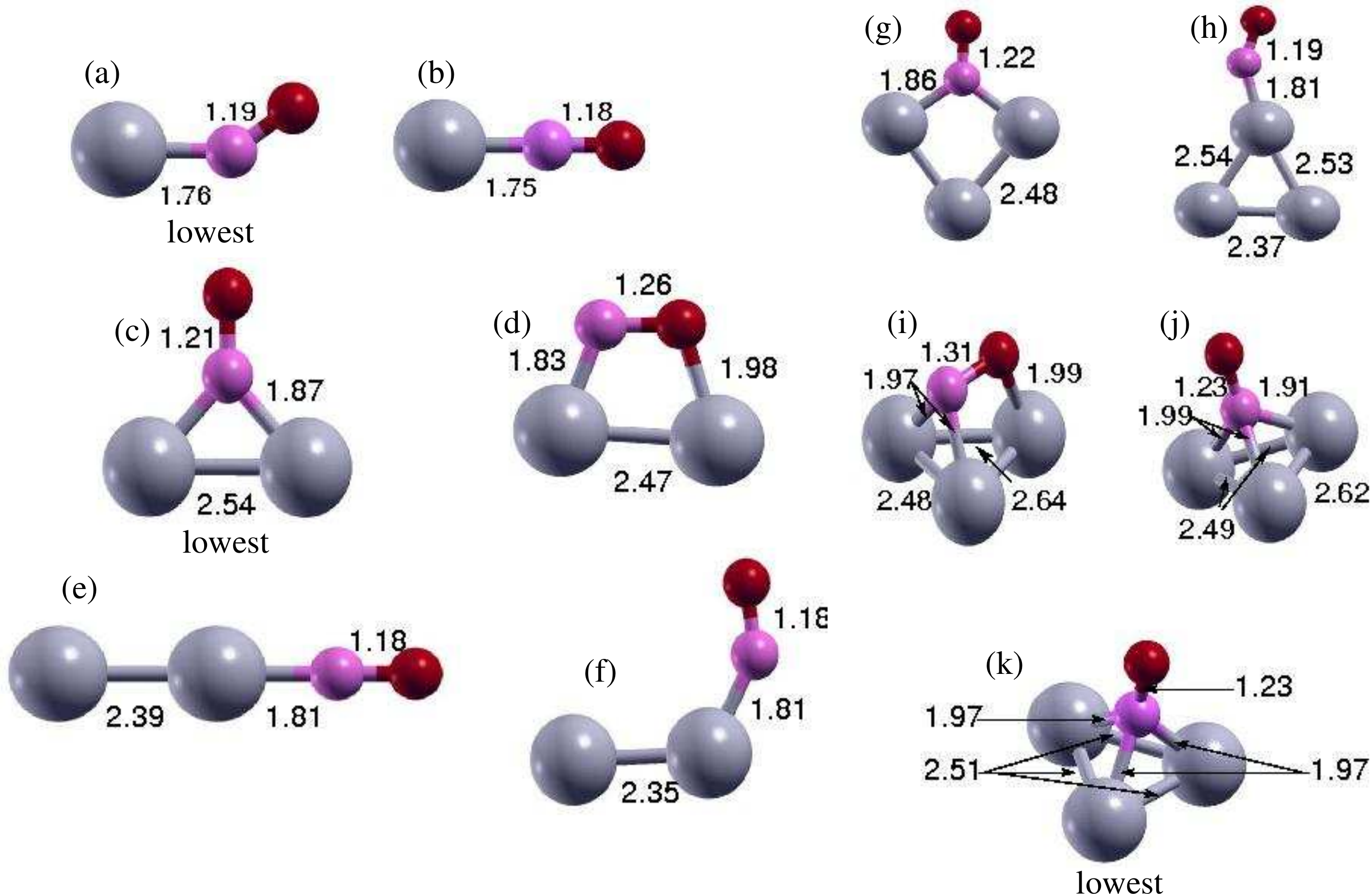}
\caption{(Color online) Stable NO adsorption geometries on Rh$_1$ [(a),(b)], Rh$_2$ [(c)-(f)]
and Rh$_3$ [(g)-(k)] after geometry optimization (for SP cases).
The Rh atoms are represented by grey spheres, N atoms by magenta spheres
and oxygen by red spheres. The same color convention has been
followed in Figs.~\ref{fig_norh4}, \ref{fig_norh5} and \ref{spindiff}.
The numbers in the figures are the bond lengths in angstroms.}
\label{fig_norh123}
\end{figure}

\begin{figure}[p]
\centering
\includegraphics[scale=0.40, angle=-90]{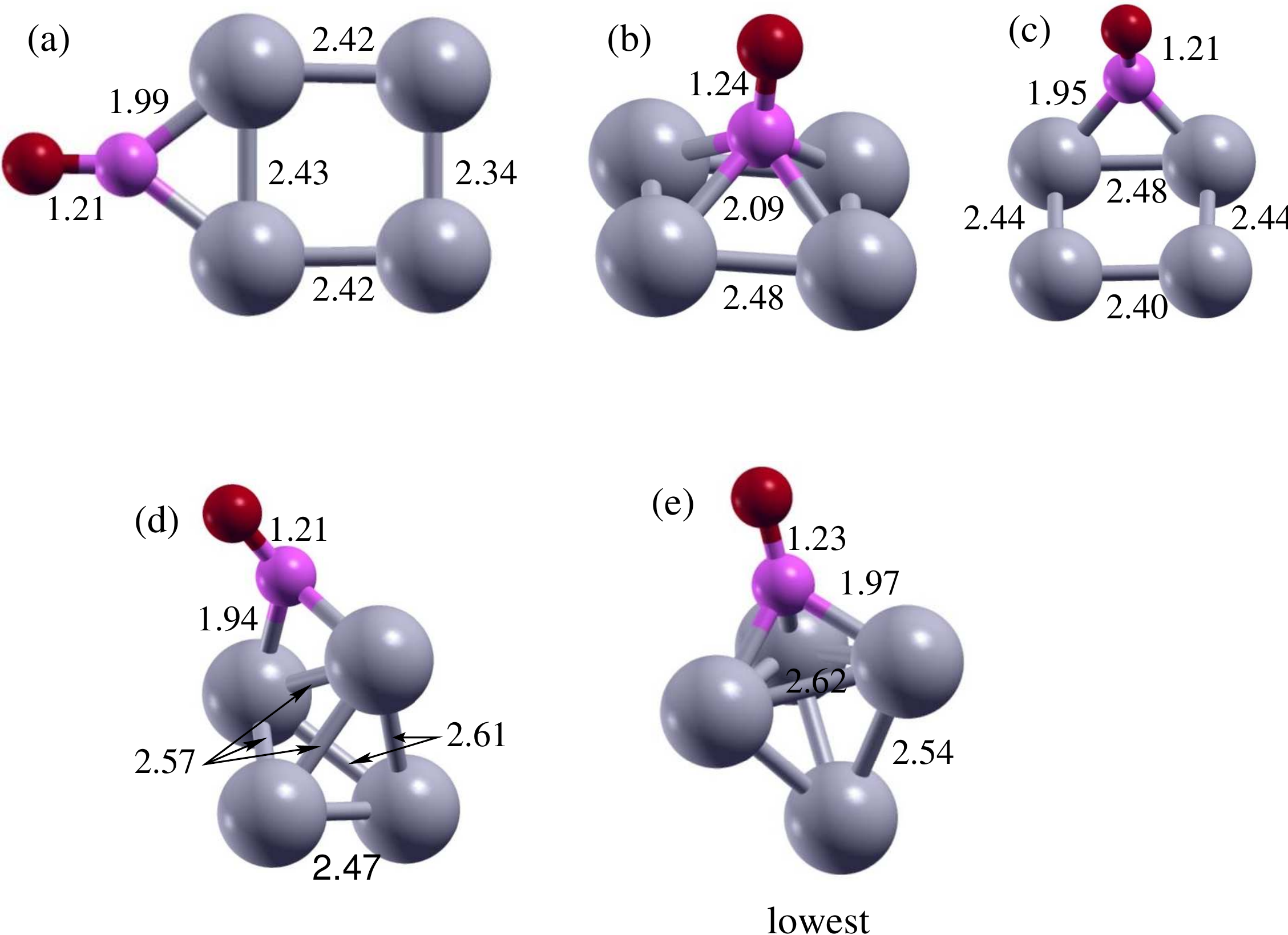}
\caption{(Color online) Stable NO adsorption geometries on Rh$_4$ after geometry optimization
(SP case). The numbers in the figures are the bond lengths in angstroms.
See caption to Fig.~\ref{fig_norh123} for color code.}
\label{fig_norh4}
\end{figure}

\begin{figure}[p]
\centering
\includegraphics[scale=0.40, angle=-90]{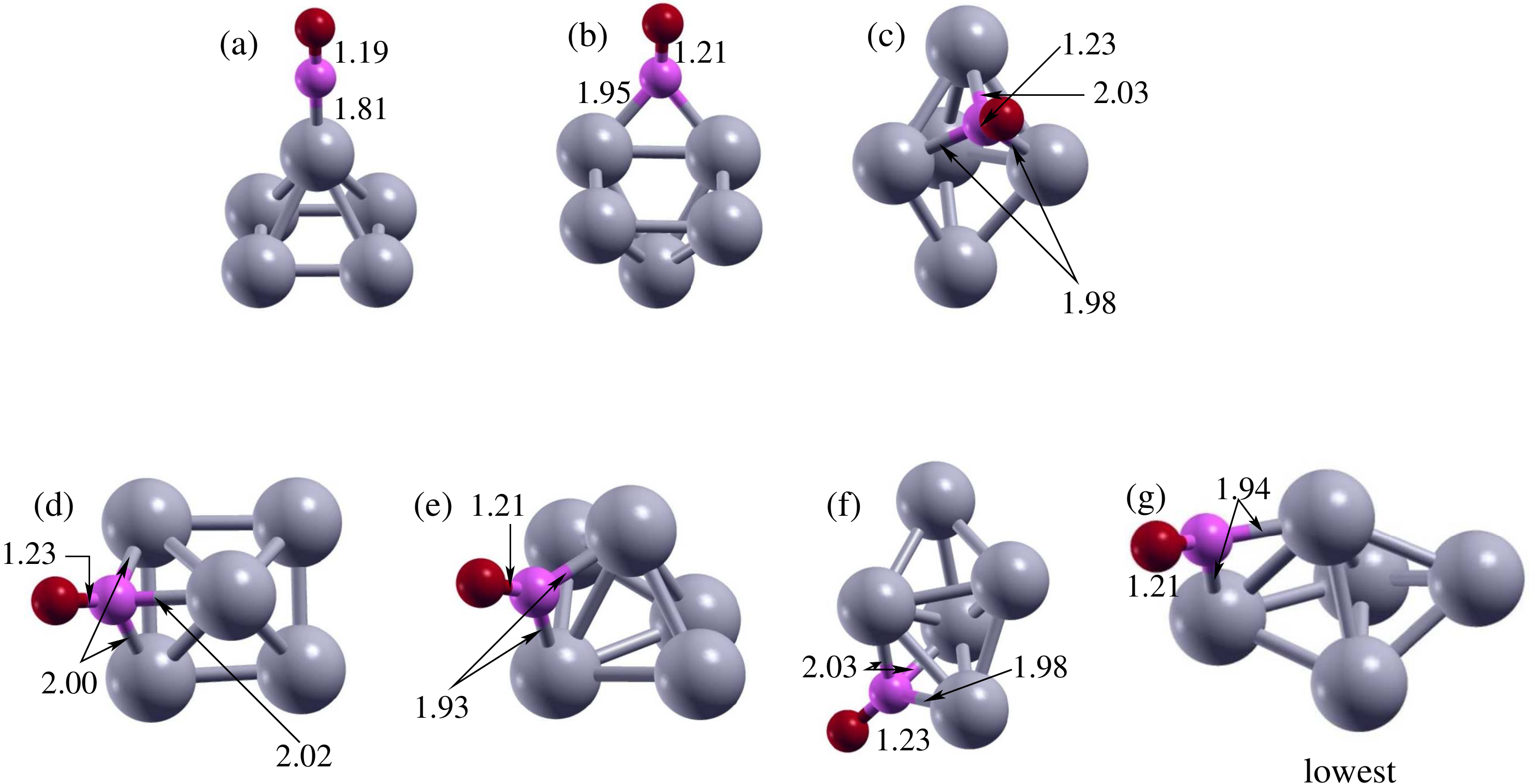}
\caption{(Color online) Stable NO adsorption geometries on Rh$_5$ after geometry optimization
(SP case). The numbers in the figures are the bond lengths in angstroms.
See caption to Fig.~\ref{fig_norh123} for color code.}
\label{fig_norh5}
\end{figure}

By considering a variety of starting spin states and geometries, we have found several stable
configurations when NO is adsorbed on the Rh clusters. The structures of many of these are
depicted in Figs.~\ref{fig_norh123}, \ref{fig_norh4} and \ref{fig_norh5}, and the results are
summarized in Table~\ref{tab_norhn_sp}, where, for purposes of comparison, we have also presented
our calculated values on the Rh(100) and Rh(111) surfaces. \cite{unpub}

 We find that on a single Rh atom, it is most favorable
for NO to adsorb in a ``bent" configuration [Fig.~\ref{fig_norh123}(a)], while on the Rh dimer,
the ``vertical bridge" configuration [Fig.~\ref{fig_norh123}(c)] is favored. On the Rh trimer,
the configuration with lowest energy corresponds to one in which NO sits, perpendicularly, on the hollow site of
an equilateral triangle [Fig.~\ref{fig_norh123}(k)].
Out of the many possible adsorption geometries for Rh$_4$NO depicted in Fig.~\ref{fig_norh4}, the one shown
in Fig.~\ref{fig_norh4}(e) is most optimal, with the nitrogen atom occupying the hollow site on one of the
triangular faces of tetrahedral Rh$_4$, and the molecule being oriented perpendicular to the face. However,
the vertical bridge configuration on the tetrahedron [Fig.~\ref{fig_norh4}(d)] is very nearly degenerate
to this. 
For the Rh$_5$NO complex, the most favored geometry is that shown in Fig.~\ref{fig_norh5}(g), where
NO sits slightly tilted on one of the short edges of the tpb. We note that when Rh$_5$ is instead in the sqp geometry, the lowest-energy configuration
corresponds to Fig.~\ref{fig_norh5}(b). Though these two configurations differ by 0.14 eV in energy, it
is possible that the latter configuration may be stabilized by kinetic barriers, and thus may be
observed in experiments.

Note that on Rh$_2$ and Rh$_5$, the most favored adsorption site is a bridge site, while on
Rh$_3$ and Rh$_4$, it is a hollow site.
We attempted to use our results to formulate guidelines for determining adsorption sites and geometries -- for example, by examining bond lengths in the bare clusters and/or looking at the ambient electron density at the adsorption site, and by comparison with
the Rh(100) surface (where NO adsorbs on a bridge site) and the Rh(111) surface (where NO adsorbs in
the hexagonal-close-packed hollow site). However, we were not able to determine any clear trends, and thus the problem of
determining NO adsorption sites on larger clusters is likely to require a systematic trial of
all possibilities, and thus considerable computational
effort.

\begin{figure}[p]
\centering
\includegraphics[scale=0.40, angle=-90]{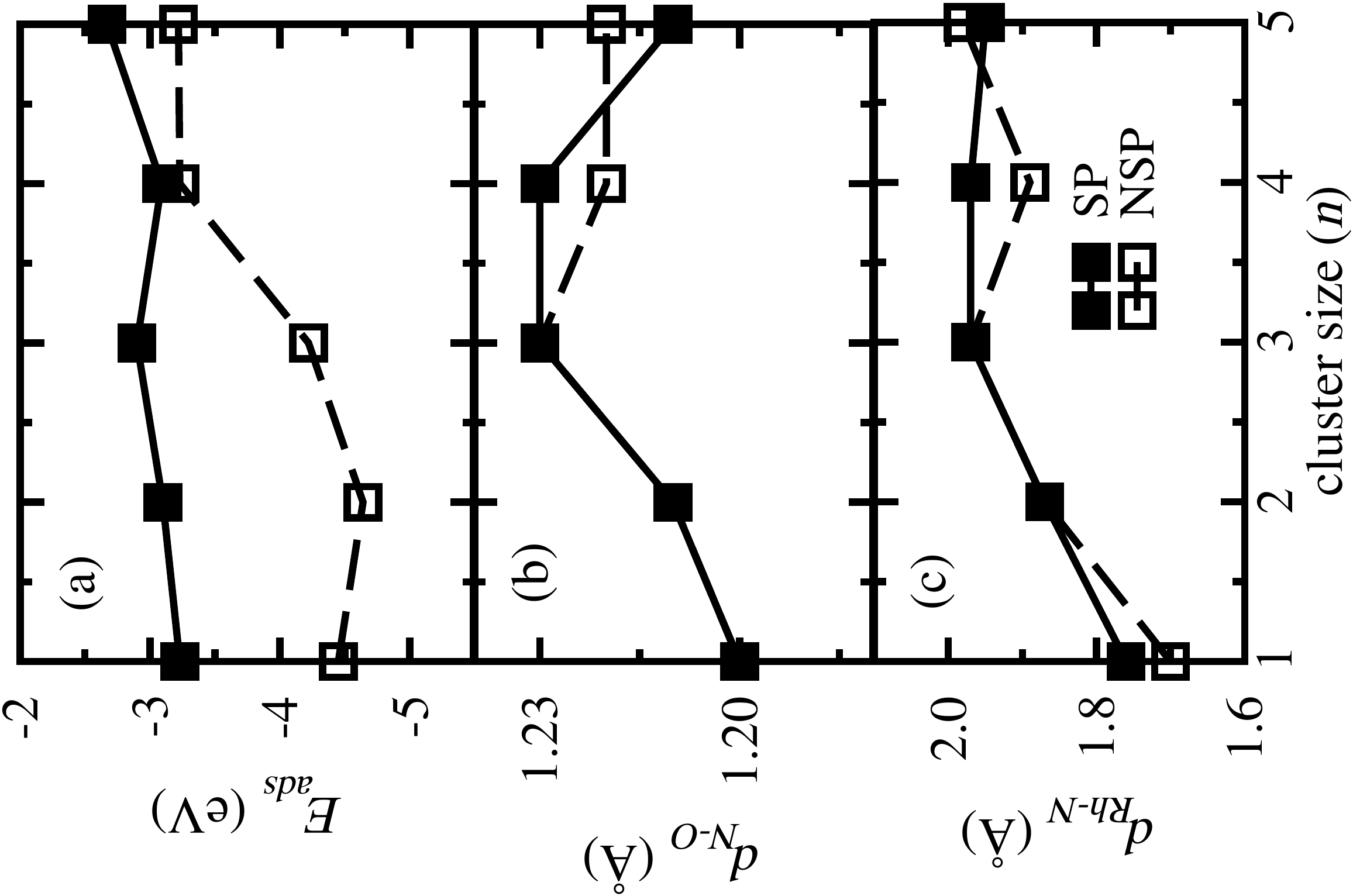}
\caption{(a) $E_{ads}$ (NO adsorption energy in eV/NO molecule), (b) N-O bond lengths ($d_{N-O}$)
and (c) Rh-N bond lengths ($d_{Rh-N}$) of Rh$_n$NO complexes
as a function of $n$, the number of atoms in the cluster.
The filled squares denote the SP results whereas the open squares denote
the NSP results.}
\label{fig_be+dsvsn}
\end{figure}

We define the adsorption energy, $E_{ads}$, by
\beq
E_{ads}=E_{Rh_nNO}-E_{Rh_n^0}-E_{NO},
\eeq

\noindent
where $E_{Rh_nNO}$ is the total energy of the Rh$_n$NO complex,
$E_{Rh_n^0}$ is the total energy of the lowest-lying isomer of the bare Rh$_n$ cluster, and $E_{NO}$ is the total
energy of the NO molecule in the gas phase. From the values listed in Table~\ref{tab_norhn_sp}, or the filled black squares in Figs.~\ref{fig_be+eadsvsneff}(b) and \ref{fig_be+dsvsn}(a), we see that $E_{ads}$ does not change significantly as a function of cluster size
and/or $\langle N_{eff} \rangle$, though its magnitude is significantly higher on the clusters than
on the flat Rh surfaces.

\begin{figure}[p]
\centering
\includegraphics[scale=0.40, angle=-90]{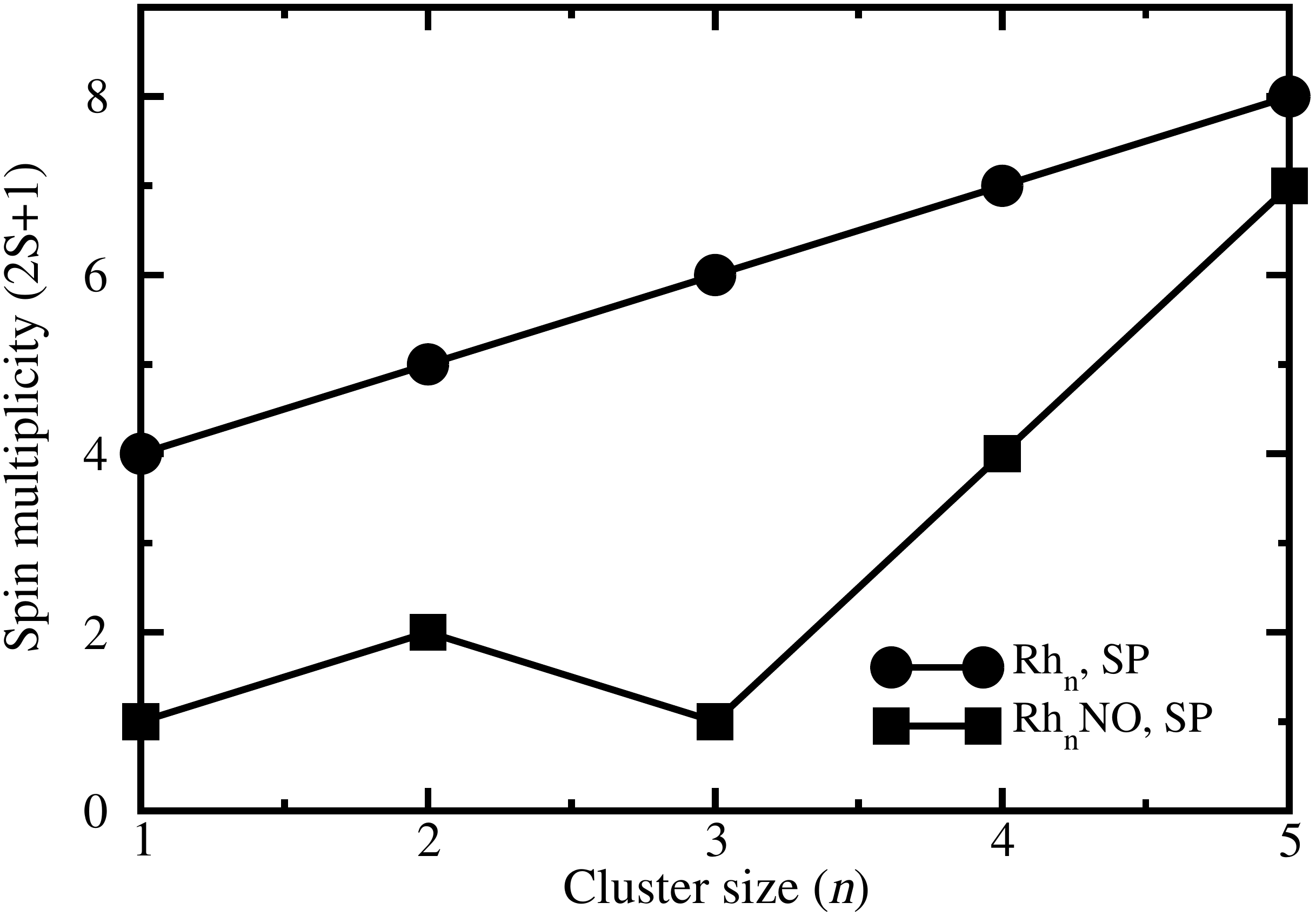}
\caption{Spin multiplicity of Rh$_n$ as a function of cluster size,
both before (filled circles) and after NO adsorption (filled squares).}
\label{spinvsn}
\end{figure}

\begin{figure}[p]
\centering
\includegraphics[scale=0.40, angle=-90]{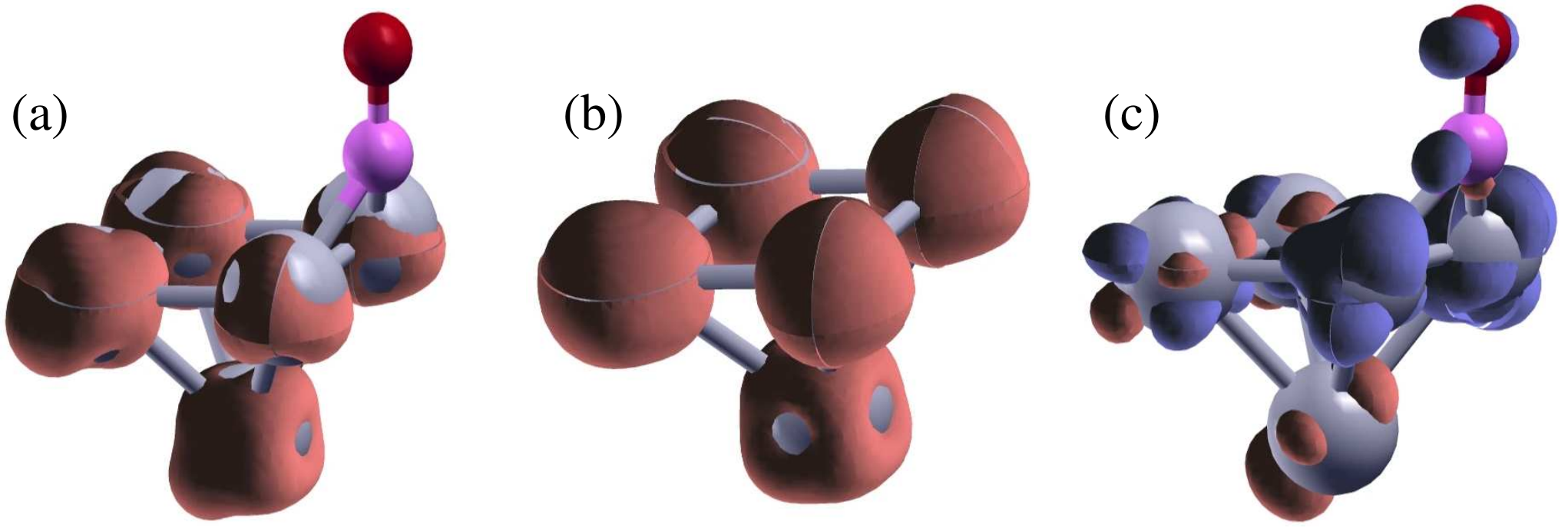}
\caption{(Color online) Differences between $\uparrow$ and $\downarrow$ charge densities ($\delta \rho^{s}$)
of (a) Rh$_5$NO and (b) Rh$_5$, but in the same geometry as in Rh$_5$NO. The difference
in $\delta \rho^{s}$ between (a) and (b) is shown in (c). The red isosurfaces in (c) denote an
increase in magnetization and the blue isosurfaces represent a decrease in magnetization. Note the quenching of magnetism in the vicinity of NO in (a),
which shows up more clearly in the blue lobes in (c).}
\label{spindiff}
\end{figure}

In Fig.~\ref{spinvsn}, we have shown how the spin multiplicity varies with $n$, for both the bare
clusters and the Rh$_n$NO complexes. It is clear that in all cases, the effect of NO adsorption is
to lower the magnetization significantly -- note that in the cases of RhNO and Rh$_3$NO, the complex
is actually found to be non-magnetic. Upon examining SP charge densities, we find that the
magnetism is quenched most strongly for the Rh atoms that are bonded to the NO molecule; the magnetization
in the vicinity of other atoms remains either essentially unchanged or, in a few cases, actually increases.
As an example, in Fig.~\ref{spindiff} we have plotted the {\it difference} between the $\uparrow$ and
$\downarrow$ densities, for (a) the lowest-energy NO-Rh$_5$ complex, and (b) the Rh$_5$ cluster, not
in its equilibrium geometry but in the same structure as in (a). A visual inspection of the two figures
shows that, in Fig.~\ref{spindiff}(a), the spin polarization is reduced significantly in the immediate neighborhood of the NO adsorption site. This becomes even clearer in Fig.~\ref{spindiff}(c), where we have plotted the difference
between (a) and (b), i.e., the change in spin polarized charge density as a result of NO adsorption. In
this figure, red and blue indicate an increase and decrease, respectively, in the degree of spin polarization. It is very clear that the five atoms of the Rh cluster fall into three groups: (i) in the
two Rh atoms bonded to NO the spin polarization decreases (by about 34\%), (ii) the next two Rh atoms 
show a redistribution of spin polarized charge density, with one set of d orbitals becoming more spin polarized
while another set becomes less spin polarized; the overall magnetization for these two atoms does not
change significantly, (iii) the remaining Rh atom, which is furthest away from NO, exhibits an
increase in spin polarization (by about 13\%); this is presumably because its bonds to the other
Rh atoms are weakened (as they are now bonded to NO), and it is only in this one atom that the
competition between magnetism and bonding is won by the former tendency.

Figs.~\ref{fig_be+dsvsn}(b) and (c) show how the N-O distance and the Rh-N distance vary with cluster size.
The variation in the former is negligible, and within the limits of accuracy of our calculations.
However, we note that, as is to be expected, the N-O bond lengths in Rh$_n$NO
are always
larger than in NO in the gas phase. The distance between the Rh and N atoms increases as $n$ increases,
possibly indicating a weaker bond between NO and the cluster. The slightly non-monotonic character
of the graph in Fig.~\ref{fig_be+dsvsn}(c) arises
from the fact that the NO adsorption site varies with $n$: in some cases it is the bridge site, while
in others it is instead the hollow site, resulting in slightly longer Rh-N bond lengths.

\subsection{NO on Rh clusters: Non-Spin-Polarized}

The results presented in the three previous sections already hint that magnetism may significantly
affect the reactivity of Rh clusters. This becomes clear when we re-do our calculations of the 
previous section for the NSP case, the results of which have been summarized in
Table \ref{fig_norhn_nsp}. When magnetism is suppressed, we find that the
magnitude of the adsorption energy
$E_{ads}$ is significantly increased, and now varies monotonically with $n$, with the adsorption being
most favored on the single Rh atom and least so on Rh$_5$ (see Fig~\ref{fig_be+dsvsn}(a)). The open squares in
Fig.~\ref{fig_be+eadsvsneff}(b) show how $E_{ads}$ varies with $\langle N_{eff} \rangle$ in the NSP case; note the monotonic 
dependence as well as the difference from the SP situation (filled black squares). It is seen that
doing an NSP calculation leads to significantly higher values in the
magnitude of the adsorption energy than those
obtained upon performing SP calculations.

Looking at the open squares in Figs.~\ref{fig_be+dsvsn}(b) and (c), we see that, in contrast to the
behavior of $E_{ads}$, the N-O and Rh-N bond lengths do not appear to be very sensitive to magnetism
(for RhNO and Rh$_3$NO, the SP and NSP results are identical, since the complexes are non-magnetic).

\section{Discussion and Summary}
\label{Disc}
We list here the main findings of the previous section: (i) small Rh clusters are magnetic, as is
well-established in the literature, (ii) the magnetic moment per atom (for the lowest-energy configuration) decreases monotonically with the size of the cluster, (iii) the average effective coordination number and average bond
length both increase as the size of the cluster increases, (iv) as a result of the magnetism of
the clusters, the bond lengths increase and effective coordination decreases, relative to what they
would be if the clusters were non-magnetic (v) the adsorption energy $E_{ads}$ does not display
any clear trend as a function of $n$ or $\langle N_{eff} \rangle$, in the SP case, (vi) however,
$E_{ads}$ is larger (in magnitude) for the clusters than on Rh(100) or Rh(111), (vii) the adsorption
of NO strongly quenches (and, in some cases, eliminates) the magnetization of the bare clusters,
(viii) this effect is local, being most prominent on the Rh atoms bonded to the adsorbate, (ix)
repeating the calculations with spin polarization suppressed leads to still higher adsorption energies.

Our finding that adsorption is weaker in the magnetic (and less effectively coordinated) case than in the non-magnetic one may seem, initally, to contradict the general expectation that lower coordination 
favors increased binding. However, this apparent contradiction arises from the fact that NO
adsorption quenches the magnetism on the Rh clusters. This is
similar to what has been observed for CO and NO adsorption
on the magnetic Ni(110) surface\cite{Jenkins1,Jenkins2,Dalcorso}, as
well as for NO adsorption on a Rh monolayer in a hypothetical
(bulk-truncated) structure.\cite{Hass} However, in the last of
these cases (Ref.~\onlinecite{Hass}) we note that the adsorption
of NO completely removed the magnetization of the Rh atoms,
whereas in our case this effect is only partial. For all the
Rh$_n$ and Rh$_n$NO cases we have examined, the lowest-energy SP solution is lower than the lowest-energy NSP solution (except for RhNO
and Rh$_4$NO, where the two are equal in energy, as the magnetization is totally quenched upon adsorption of NO). However, the difference in total energy between the NSP and SP Rh$_n$ clusters is much larger
than the difference in the total energies of the NSP and SP Rh$_n$NO complexes, due to the quenching
of magnetism in the latter. As a direct consequence of this, the SP adsorption energies are reduced
with respect to the NSP adsorption energies.
Our findings are consistent with those of Nayak {\it et al.},\cite{Nayak} who studied Rh$_n$H$_2$
clusters, and found that H$_2$ binds much more strongly to a non-magnetic isomer of Rh$_4$ than to
a magnetic one.

Thus, lower coordination (which favors binding to adsorbates) also favors magnetism (which disfavors
binding to adsorbates). This suggests that for Rh nanocatalysis, there may be an optimally effective
cluster size, which is low enough to favor increased binding, while being high enough so that magnetism
does not significantly reduce binding. The next step would be to see whether the greater binding to
Rh clusters (relative to flat Rh surfaces) will also result in lower barriers for NO dissociation;
work in this direction is in progress.

We note that adsorption studies such as the ones presented here could conceivably yield three possible
indicators of dissociation barriers: (i) the intramolecular (N-O) bond length (ii) the metal-adsorbate
(Rh-N) bond length (iii) the adsorption energy $E_{ads}$. One would expect that lowering of dissociation barriers might correlate with an increase in
magnitude of (i) and (iii), and a decrease in (ii). Of these, it seems
likely that (i) will not be a very reliable indicator, given that the N-O distance is almost the same
on Rh(100) and Rh(111), even though the dissociation barriers on the
two surfaces are different. Indeed,
our results show almost no size-dependence for $d_{N-O}$, in both SP and NSP cases. We find that $d_{Rh-N}$
decreases as $n$ is decreased, possibly indicating a stronger Rh-N and thus weaker N-O bond. However,
the values of $E_{ads}$ do not show such a size-dependent trend for the SP
case. Thus, calculations of dissociation
barriers are needed to resolve the issue of whether or not the barriers will vary significantly as a
function of size.

In conclusion, our results illustrate the fact that in small clusters there can be a competition between the tendencies to bond and to magnetize, with both effects being favored by reduced coordination. This interplay between bonding and magnetism has a significant influence on the size-dependent trends in
the chemical behavior of such systems.

\section{Acknowledgments}

We thank Stefano Baroni for encouraging this collaboration, which was made
possibly by a grant from the Indo-Italian Program of Cooperation in Science
and Technology 2005-2007, funded by the Department of Science and Technology,
Government of India, and the Ministero degli Affari Esteri, Government of
Italy. Helpful conversations with Erio Tosatti are acknowledged. 
We thank Dr. Leonardo Gastaldi, Dr. Sadhana Relia and Jimmy Alexander for their assistance.
P.G. also acknowledges CSIR, India, for a research scholarship, and S.N.
acknowledges the Associateship Programme of the Abdus Salam International
Centre for Theoretical Physics.

\pagebreak
\newpage
\clearpage


\begin{table}
\caption{Comparison of our results (spin-polarized) on Rh$_n$
with previous calculation and experiments. The abbreviations for the different
geometries of the bare clusters have been explained in Section~\ref{FPdet}. The
numbers in parentheses against the bond lengths indicate the number of
such bonds. For a single Rh atom in the gas phase, the spin multiplicity is 4.}
\begin{center}
\begin{tabular}{cccccccc}
\hline
\hline
System&Author&Geometry&$\langle N_{eff}\rangle$&BE &Bond lengths&Spin multiplicity\\
&&&&(eV/atom)&(\AA)&(2S+1)\\
\hline
&Present&-&3.61&1.48&2.25&5\\
Rh$_2$& Ref.~\onlinecite{Nayak}&-&-&1.51&2.26&5\\
&Ref~\onlinecite{Reddy2}&-&-&1.88&2.34&5\\
&Ref~\onlinecite{Gingerich}&-&-&1.46&2.28&5\\
\hline
&Present&eq & 4.75&2.03&2.46&6\\
&Present&eq &4.67&2.00&2.41&4\\
&Present&isos & 4.14&2.04 & 2.56, 2.41 ($\times$2)&6\\
Rh$_3$&Ref~\onlinecite{Nayak}&eq&-&1.99& 2.42&4\\
&Ref~\onlinecite{Nayak}&isos&-&1.94&2.53 ($\times$2), 2.40 &6\\
&Ref~\onlinecite{Reddy2}&eq&-&2.35& 2.45&4\\
&Ref~\onlinecite{Reddy2}&isos&-&2.37&2.52 ($\times$2), 2.48 &6\\
\hline
Rh$_4$ & Present & sq & 5.34 & 2.37 & 2.36 & 5\\
& Present & tet & 5.04 & 2.42 & 2.53 & 7\\
& Ref~\onlinecite{Endou1} & tet & -& 3.38 & 2.40 & 7\\
& Ref~\onlinecite{Nayak} & tet & -& 2.41 & 2.49 & 1\\
& Ref~\onlinecite{Reddy2}& tet & -& 2.91 & 2.50 & 1\\
\hline
Rh$_5$ & Present & sqp & 5.88 & 2.74 & 2.45 (sq base), 2.55 &  6\\
       & Present & sqp & 5.28 & 2.74 & 2.44 (sq base), 2.62 &  8\\
 & Present & tbp & 5.57 & 2.71 & 2.66 ($\times$3), 2.52 ($\times$6) & 8\\
 & Ref~\onlinecite{Reddy2}& sqp & - & 3.13 & 2.48 (sq base),
   2.63 & 8\\
 & Ref~\onlinecite{Futschek} & sqp & - & 3.03 & & 6\\
 & Ref~\onlinecite{Reddy2}& tbp & -& 3.11 & 2.57 ($\times$3), 2.46 ($\times$6) & 6\\
 & Ref~\onlinecite{Futschek} & tbp &-& 2.97 & & 8\\
\hline
\end{tabular}
\end{center}
\label{tab_bare_sp}
\end{table}


\pagebreak
\newpage
\clearpage


\begin{table}
\caption{Geometry, effective coordination number ($\langle N_{eff}\rangle$), binding energy (BE) and Rh-Rh bond length ($d_{Rh-Rh}$) of
Rh$_n$ (non spin-polarized). The numbers in the parentheses indicate
the number of such Rh-Rh bonds present in the clusters.
The abbreviations for the different
geometries of the bare clusters have been explained in Section~\ref{FPdet}.}
\begin{center}
\begin{tabular}{ccccc}
\hline
\hline
System&Geometry&$\langle N_{eff}\rangle$&BE&$d_{Rh-Rh}$\\
&&&(eV/atom) & (\AA) \\
\hline
Rh$_2$ & - & 4.38 & 0.80 & 2.18\\
Rh$_3$ & eq & 4.74 & 1.71 & 2.36\\
Rh$_4$ & sq &5.66  &2.24&2.34\\
       & tet & 5.93 & 2.37 & 2.47 \\
Rh$_5$ & tbp & 6.26 & 2.54 & 2.49 ($\times$3), 2.53 ($\times$6)\\
       & sqp & 6.15 & 2.59 & 2.43 (sq base), 2.54 \\
\hline
\end{tabular}
\end{center}
\label{tab_bare_nsp}
\end{table}


\pagebreak
\newpage
\clearpage

\begin{table*}
\caption{Geometry, NO adsorption energy ($E_{ads}$), Rh-N bond length ($d_{Rh-N}$),
N-O bond length ($d_{N-O}$) and spin multiplicity of Rh
clusters with NO adsorbed on them. The lowest energy configuration 
for each system is given in bold text. For the sake of comparison
we also present data on the Rh(100) and Rh(111) surfaces.}
\begin{center}
\begin{tabular*}{6.0in}{c@{\extracolsep{\fill}}c@{\extracolsep{\fill}}
c@{\extracolsep{\fill}}c@{\extracolsep{\fill}}c@{\extracolsep{\fill}}
c@{\extracolsep{\fill}}}
\hline
\hline
System & geometry &$E_{ads}$ & $d_{Rh-N}$ & $d_{N-O}$ &
Spin multiplicity\\
 & & eV/NO molecule &(\AA) & (\AA) & (2S+1) \\
\hline
RhNO & \textbf{Fig.~\ref{fig_norh123}(a)} & \textbf{-3.23} & \textbf{1.76} & \textbf{1.20} &
\textbf{1} \\
  & Fig.~\ref{fig_norh123}(b) & -3.17 & 1.75 & 1.18 & 1 \\
\hline
Rh$_2$NO & \textbf{Fig.~\ref{fig_norh123}(c)} & \textbf{-3.10} & \textbf{1.87} &
\textbf{1.21} & \textbf{2} \\
   & Fig.~\ref{fig_norh123}(d) & -2.67 & 1.83 & 1.26 & 2\\
   & Fig.~\ref{fig_norh123}(e) & 2.22 & -1.81 & 1.18 &4 \\
   & Fig.~\ref{fig_norh123}(f) & -2.46 & 1.81 & 1.18 &4\\
\hline
Rh$_3$NO & Fig.~\ref{fig_norh123}(g) & -2.80 & 1.86 & 1.22 & 3 \\
         & Fig.~\ref{fig_norh123}(h) & -2.71 & 1.81 & 1.19 & 5 \\
         & Fig.~\ref{fig_norh123}(i) & -2.21 & 1.97 & 1.31 & 3 \\
         & Fig.~\ref{fig_norh123}(j) & -2.89 & 1.91, 1.99 & 1.23 & 3\\
         & \textbf{Fig.~\ref{fig_norh123}(k)} & \textbf{-2.90} & \textbf{1.97} &
\textbf{1.23} & \textbf{1}\\
\hline
Rh$_4$NO& Fig.~\ref{fig_norh4}(a) & -2.09 & 1.99 & 1.21 & 6 \\
        & Fig.~\ref{fig_norh4}(b) & -1.69 & 2.09 & 1.24 & 4 \\
        & Fig.~\ref{fig_norh4}(c) & -2.79 & 1.95 & 1.21 & 6 \\
        & Fig.~\ref{fig_norh4}(d) & -3.08 & 1.21 & 1.24 & 6 \\
        & \textbf{Fig.~\ref{fig_norh4}(e)}  & \textbf{-3.09} & \textbf{1.97} & \textbf{1.23} &
\textbf{4} \\
\hline
Rh$_5$NO & Fig.~\ref{fig_norh5}(a) &-2.57 & 1.81 & 1.19 & 3 \\
         & Fig.~\ref{fig_norh5}(b) & -2.67 & 1.95 & 1.21 & 7 \\
         & Fig.~\ref{fig_norh5}(c) & -2.50 & 1.98, 2.03 & 1.23 & 7 \\
         & Fig.~\ref{fig_norh5}(d) & -2.48 & 2.00, 2.02 & 1.23 & 7 \\
         & Fig.~\ref{fig_norh5}(e) & -2.42 & 1.93 & 1.21 & 7 \\
         & Fig.~\ref{fig_norh5}(f) & -2.79 & 1.98, 2.03 & 1.23 & 7 \\
         & \textbf{Fig.~\ref{fig_norh5}(g)} & \textbf{-2.81} & \textbf{1.94} & \textbf{1.21}
 & \textbf{7} \\
\hline
Rh(100) &\textbf{vertical bridge} & \textbf{-2.59} & \textbf{1.96} &
\textbf{1.20} & - \\
        & horizontal hollow & -2.47 & 1.98 & 1.31 &- \\
\hline
Rh(111) & \textbf{hollow hcp} & \textbf{-2.18} & \textbf{2.10} &
\textbf{1.20} &-\\
\hline
\end{tabular*}
\end{center}
\label{tab_norhn_sp}
\end{table*}


\pagebreak
\newpage
\clearpage

\begin{table*}
\caption{Geometry,
adsorption energy of NO ($E_{ads}$), Rh-N bond length ($d_{Rh-N}$) and
N-O bond length ($d_{N-O}$) in Rh$_n$NO in the non-spin-polarized case.
The bond topology is similar to those indicated in the second column;
however, note that these figures represent the SP and not NSP
case. The bond lengths do differ in the two cases.}
\begin{center}
\begin{tabular*}{6.0in}{c@{\extracolsep{\fill}}c@{\extracolsep{\fill}}
c@{\extracolsep{\fill}}c@{\extracolsep{\fill}}c@{\extracolsep{\fill}}}
\hline
\hline
System & geometry &$E_{ads}$ & $d_{Rh-N}$ & $d_{N-O}$ \\
 & & (eV/NO molecule) &(\AA) & (\AA) \\
\hline
RhNO & $\sim$Fig.~\ref{fig_norh123}(a) & -4.45 & 1.70 & 1.20\\
     & $\sim$Fig.~\ref{fig_norh123}(b) & -4.39 & 1.75 & 1.18\\
\hline
Rh$_2$NO & $\sim$Fig.~\ref{fig_norh123}(c) & -4.64 & 1.87 & 1.21\\
\hline
Rh$_3$NO & $\sim$Fig.~\ref{fig_norh123}(k) & -4.22 & 1.97 & 1.23\\
\hline
Rh$_4$NO & $\sim$Fig.~\ref{fig_norh4}(d) & -3.23 & 1.89 & 1.22\\
         & $\sim$Fig.~\ref{fig_norh4}(e) & -3.07 & 1.96 & 1.24\\
\hline
Rh$_5$NO & $\sim$Fig.~\ref{fig_norh5}(g) & -3.31 & 1.94, 1.90 & 1.22\\
         & $\sim$Fig.~\ref{fig_norh5}(b) & -3.22 & 1.98 & 1.22\\
\hline
\end{tabular*}
\end{center}
\label{fig_norhn_nsp}
\end{table*}

\pagebreak
\newpage
\clearpage

\pagebreak
\newpage
\clearpage

\pagebreak
\newpage
\clearpage

\pagebreak
\newpage
\clearpage

\pagebreak
\newpage
\clearpage

\pagebreak
\newpage
\clearpage

\pagebreak
\newpage
\clearpage

\pagebreak
\newpage
\clearpage

\pagebreak
\newpage
\clearpage

\begin{references}

\bibitem{Loffreda1} D. Loffreda, D. Simon and P. Sautet, J. Chem. Phys.
\textbf{108}, 6447 (1998).

\bibitem{Borg} H. J. Borg, J. F. C.-J. M. Reijerse, R. A. van Santen, and J. W.
Niemantsverdriet, J. Chem. Phys. \textbf{101}, 10052 (1994).

\bibitem{Hopstaken} M. J. P. Hopstaken and J. W. Niemantsverdriet,
J. Vac. Sci. and Tech. A, \textbf{18}, 1503 (2000).

\bibitem{Villarubia} J.S. Villarrubia and W. Ho, J. Chem. Phys. {\bf 87},
750 (1987).

\bibitem{Zambelli} T. Zambelli, J. Wintterlin, H. Trost and G. Ertl,
Science {\bf 273}, 1688 (1996).

\bibitem{Hammer} B. Hammer, Phys. Rev. Lett. {\bf 83}, 3681 (1999).

\bibitem{Reddy1} B. V. Reddy, S. N. Khanna and B. I. Dunlap, Phys. Rev. Lett.
\textbf{70}, 3323 (1993).

\bibitem{Cox1} A. J. Cox, J. G. Louderback, and L. A. Bloomfield,
Phys. Rev. Lett. \textbf{71}, 923 (1993).

\bibitem{Cox2} A. J. Cox, J. G. Louderback, S. E. Apsel and L. A. Bloomfield,
Phys. Rev. B \textbf{49}, 12295 (1993).

\bibitem{Sanchez} A. Sanchez, S. Abbet, U. Heiz, W.-D. Schneider,
H. H\"{a}kkinen, R. N. Barnett and U. Landman, J. Phys. Chem. A \textbf{103}
9573 (199). 

\bibitem{Hannu} H. H\"{a}kkinen, S. Abbet, A. Sanchez, U. Heiz and
U. Landman, Angew. Chem. Int. Ed. \textbf{42}, 1297 (2003).

\bibitem{Yoon} B. Yoon, H. H\"{a}kkinen, U. Landman, A. S. W\"{o}rz, J.-M.
Antonietti,  S. Abbet, K. Judai and U. Heiz, Science \textbf{307},
403, (2005).

\bibitem{Walter} M. Walter and H. H\"{a}kkinen, Phys. Rev. B \textbf{72},
205440 (2005). 

\bibitem{Heiz} U. Heiz, A. Sanchez, S. Abbet and W.-D. Schneider, Chem. Phys.
{\bf 262}, 189 (2000).

\bibitem{Xu} Z. Xu, F.-S. Xiao, S.K. Purnell, O. Alexeev, S. Kawi,
S.E. Deutsch and B.C. Gates, Nature {\bf 372}, 346 (1994).

\bibitem{Abbet} S. Abbet, A. Sanchez, U. Heiz, W.-D. Schneider, A. M. Ferrari,
G. Pacchiono and N. Rosch, J. Am. Chem. Soc. \textbf{122}, 3453 (2000).

\bibitem{Bae1} Y.-C Bae, H. Osanai, V. Kumar and Y. Kawazoe, Phys. Rev. B
\textbf{70}, 195413 (2004).

\bibitem{Bae2} Y.-C Bae, V. Kumar, H. Osanai and Y. Kawazoe, Phys. Rev. B
\textbf{72}, 125427 (2005).

\bibitem{Chang} C. M. Chang and M. Y. Chou, Phys. Rev. Lett.
\textbf{93}, 133401 (2004).

\bibitem{Rogan} J. Rogan, G. Garcia, C. Loyola, W. Orellana, R. Ramirez
and M. Kiwi, J. Chem. Phys. \textbf{125}, 214708 (2006).

\bibitem{Futschek} T. Futschek, M. Marsman and J. Hafner, J. Phys. Condens.
Matt. \textbf{17}, 5927 (2005).

\bibitem{Reddy2} B. V. Reddy, S. K. Nayak, S. N. Khanna, B. K. Rao and
P. Jena, Phys. Rev. B \textbf{59}, 5214 (1999).

\bibitem{Nayak} S. K. Nayak, S. E. Weber, P. Jena, K. Wildberger, R. Zeller,
P. H. Dederichs, V. S. Stepanyuk and W. Hergert, Phys. Rev. B
\textbf{56}, 8849 (1997).

\bibitem{Chien} C.-H. Chien, E. B.-Barojas and M. R. Pederson,
Phys. Rev. B \textbf{56}, 2196 (1998).

\bibitem{Endou1} A. Endou, N. Ohashi, K. Yoshizawa, S. Takami, M. Kubo,
A. Miyamoto and E. Broclawik, J. Phys. Chem. B \textbf{104},
5110 (2000).

\bibitem{Jinlong} Y. Jinlong, F. Toigo and W. Kelin, Phys. Rev. B
\textbf{50}, 7915 (1994).

\bibitem{Harding} D. Harding, S. R. Mackenzie and T. R. Walsh,
J. Phys. Chem. B \textbf{110}, 18272 (2006).

\bibitem{Gingerich} K. A. Gingerich and D. L. Cocke, J. Chem. Soc. Chem.
Commun. \textbf{1}, 536 (1972).

\bibitem{Endou2} A. Endou, R. Yamauchi, M. Kubo, A. Stirling and
A. Miyamoto, App. Surf. Sc. \textbf{119}, 318 (1997).


\bibitem{Ford} M. S. Ford, M. L. Anderson, M. Barrow, D. P. Woodruff,
T. Drewello, P. J. Derrick, S. R. Mackenzie, Phys. Chem. Chem. Phys.
\textbf{7}, 975 (2005).

\bibitem{Anderson} M. L. Anderson, M. S. Ford, P. J. Derrick,
T. Drewello, D. P. Woodruff and S. R. Mackenzie, J. Phys. Chem. A
\textbf{110}, 10992 (2006).

\bibitem{Loffreda3} D. Loffreda, F. Delbecq, D. Simon, and P. Sautet,
J. Chem. Phys. \textbf{115}, 8101 (2001).
 
\bibitem{qnesp} http://www.quantum-espresso.org

\bibitem{Kohnsham} W. Kohn and L.J. Sham, Phys. Rev. {\bf 140}, A1133 (1965).

\bibitem{Vanderbilt} D. Vanderbilt, Phys. Rev. B {\bf 41}, 7892 (1990).

\bibitem{GGA4} J.P. Perdew, K. Burke, and M.Ernzerhof, Phys. Rev. Lett.
{\bf 77}, 3865 (1996).

\bibitem{Rhbexpt} P. Villars and L. D. Calvert, \textit{Pearson's Handbook of
Crystallographic Data for Intermetallic Phases}, American Society for Metals,
Metals Park, OH, 1985.

\bibitem{Johnson} B.G. Johnson, P.M.W. Gill, and J.A. Pople,
J. Chem. Phys. \textbf{98}, 5612 (1993).

\bibitem{Hellmann} H. Hellmann, EinfÃ¼hrung in die Quantenchemie
(Franz Deuticke, Leipzig, 1937), Sec. 54.

\bibitem{Feynman} R. P. Feynman, Phys. Rev. \textbf{56}, 340 (1939). 

\bibitem{BFGS} (a) C. G. Broyden, J. Inst. Math Appl. \textbf{6}, 76 (1970),
(b) R. Fletcher, J. Comput. \textbf{13}, 317 (1970), (c) D. Goldfarb, Math.
Comp. \textbf{24}, 23 (1970), (d) D. F. Shanno, Math. Comp. \textbf{24}, 647
(1970).

\bibitem{corelev} A. Baraldi, L. Bianchettin, E. Vesselli, S. de Gironcoli,
S. Lizzit, L. Petaccia, G. Zampieri, G. Comelli and R. Rosei,
N. J. Phys. \textbf{9}, 143 (2007).

\bibitem{Daw} M. S. Daw, S. M. Foiles, and M. I. Baskes, Mater. Sci. Rep.
\textbf{9}, 251 (1993). 

\bibitem{EMT} K.W. Jacobsen, J.K. Norskov, and M.J. Puska,
Phys. Rev. B {\bf 35}, 7423 (1987).

\bibitem{unpub} R. Pushpa, P. Ghosh, S. Narasimhan, and
S. de Gironcoli (unpublished).

\bibitem{Jenkins1} Q. Ge, S.J. Jenkins, and D.A. King, Chem. Phys. Lett.
{\bf 327}, 125 (2000).

\bibitem{Jenkins2} S.J. Jenkins, Q. Ge, and D.A. King, Phys. Rev. B {\bf 64},
012413 (2001).

\bibitem{Dalcorso} F. Favot, A. Dal Corso, and A. Baldereschi, Phys. Rev. B
{\bf 63}, 115416 (2001).

\bibitem{Hass} K. C. Hass, M.-H. Tsai and R. V. Kasowski, Phys. Rev. B
\textbf{53}, 44 (1996).

\end{references}
\end{document}